 \documentclass[
a4paper,showpacs,preprint, amsmath, amssymb, floatfix,
nofootinbib,wrapfloat byrevtex]{revtex4}

\usepackage{lineno,hyperref}
\modulolinenumbers[5]

 \usepackage{amsmath,mathrsfs,graphicx,epstopdf,placeins,float}
 \usepackage{booktabs}
 \usepackage{multirow}
 \usepackage[titletoc]{appendix}
 \usepackage{titlesec}

 \newcommand{\beq}[1]{\begin{equation}\label{#1}}
 \newcommand{\eeq}{\end{equation}}
 \newcommand{\bea}[1]{\begin{eqnarray}\label{#1}}
 \newcommand{\eea}{\end{eqnarray}}
 \newcommand\figcaption{\def\@captype{figure}\caption}
 \newcommand\tabcaption{\def\@captype{table}\caption}

\begin{document}
 \title{Supersymmetric quantum mechanics and topology}
 \author{Muhammad Abdul Wasay $^{\!\!1}$}
\email{wasay\_31@hotmail.com}
\affiliation{$^1$Department of Physics, University of Agriculture, Faisalabad 38040, Pakistan}
 \begin{abstract}
    Supersymmetric quantum mechanical models are computed by the Path integral approach. In the $\beta\rightarrow0$ limit, the integrals localize to the zero modes. This allows us to perform the index computations exactly because of supersymmetric localization, and we will show how the geometry of target space enters the physics of sigma models resulting in the relationship between the supersymmetric model and the geometry of the target space in the form of topological invariants. Explicit computation details are given for the Euler characteristics of the target manifold, and the index of Dirac operator for the model on a spin manifold.
 \end{abstract}
 \maketitle
 \smallskip

\section{Introduction}
 \vspace{2mm}
 Supersymmetry is a quantum mechanical space-time symmetry which induce transformations between bosons and fermions. The generators of this symmetry are spinors which are anticommuting (fermionic) variables rather than the ordinary commuting (bosonic) variables, hence their algebra involves anticommutators instead of commutators. A unified framework consisting of both bosons and fermions thus became possible, both combined in the same supersymmetric multiplet\cite{zumino}. It is overwhelmingly accepted that supersymmetry is an essential feature of any unified theory as it not only provides a unified ground for bosons and fermions but is also helpful in reducing ultraviolet divergences. It was discovered by Gel'fand and Likhtman\cite{Likhtman}, Ramond\cite{Ramond} and Neveu and Schwarz\cite{Schwarz} and later by a few physicists\cite{volkov}\cite{zumino}. Whether Supersymmetry (SUSY) is actually realized in nature or not is still not clear however it has provided powerful mathematical tools and enormous amount of insights have been obtained\cite{greene}. For example, SUSY could be used to unify the space-time and internal symmetries of the S-matrix avoiding the no-go theorem of Coleman and Mandula\cite{mandula}, imposing local gauge invariance to SUSY gives rise to supergravity\cite{freedman}\cite{deser}. In such theories, locally gauged SUSY gives rise to Einstein's general theory of relativity, which highlights that the local SUSY theories give a natural framework for the unification of gravity and other fundamental forces.

    Supersymmetric quantum mechanics was originally developed by Witten\cite{witten}, as a toy model to test the breaking of supersymmetry. In answering the same question, SUSY was also studied in the simplest case of SUSY QM by Cooper and Freedman\cite{cooper}. In a later paper, the so called ``Witten Index" was proposed by Witten\cite{witten1}, which is a topological invariant and it essentially provides a tool to study the SUSY breaking non-perturbatively. A year later, Bender, Cooper and Das\cite{bender} proposed a new critical index to study SUSY breaking in a lattice regulated system non-perturbatively. In its early days, SUSY QM was studied as a test to check the SUSY breaking non-perturbatively.

   Later, when people started to explore further aspects of SUSY QM, It was realized that this was a field of research worthy of further exploration in its own right. The introduction of the topological index by Witten\cite{witten1}, attracted a lot of attention from the physics community and people started to study  different topological aspects of SUSY QM.

   Witten index was extensively explored and it was shown that the index exhibited anomalies in certain theories with discrete and continuous spectra\cite{comtet}\cite{khare}\cite{cecotti}\cite{cooper1}\cite{Salomonson}. Using SUSY QM, proofs of Atiyah-Singer Index theorem was given\cite{Gaume}\cite{Gaume1}\cite{Gaume2}. A link between SUSY QM and stochastic differential equations was investigated in\cite{parisi}, which was used to prove algorithms about stochastic quantization, Salomonson and Van Holten were the first to give a path integral formulation of SUSY QM\cite{holten}. The ideas from SUSY QM were extended to study higher dimensional systems and systems with many particles to implement such ideas to problems in different branches of physics e.g., condensed matter physics, atomic physics and statistical physics etc.\cite{prl}\cite{khare1}\cite{prl1}\cite{prl2}\cite{prl3}\cite{stedman}. Another interesting application is\cite{gauntlett}, which is that the low energy dynamics of $k$-monopoles in $\mathcal{N}=2$ supersymmetric Yang-Mills theory are determined by an $\mathcal{N}=4$ supersymmetric quantum mechanics based on the moduli space of $k$ static monopole solutions.

   There are also situations where SUSY QM arises naturally, e.g., in the semi-classical quantization of instanton solitons in field theory. In the classical limit, the dynamics can often be described in the terms of motion on the moduli space of the instanton solitons. Semi-classical effects are then described by quantum mechanics on the moduli space. In a supersymmetric theory, soliton solutions generally preserve half the supersymmetries of the parent theory and these are inherited by the quantum mechanical system. Complying with this, Hollowood and Kingaby in \cite{hollowood} show that a simple modification of SUSY QM involving the mass term for half the fermions naturally leads to a derivation of the integral formula for the $\chi_y$ genus, which is a quantity that interpolated between the Euler characteristic and arithmetic genus.

     The research work in the direction of using supersymmetry to exploit topology occurred in phases, first one started in early 80s with the work of Witten\cite{witten}\cite{witten2}, Alvarez Gaume\cite{gaume3}, Friedan and Windey\cite{windey} and the later phase starting from late 80s and early nineties is still going on. A couple of major breakthroughs in the second phase were due to Witten, in \cite{witten3}, Jone's polynomials for knot invariants were understood quantum field theoretically, and in\cite{witten4}, Donaldson's invariants for four manifolds. Supersymmetric localization is a powerful technique to achieve exact results in quantum field theories. A recent development using supersymmetric localization technique is the exact computation of the entropy of $AdS_4$ black holes by a topologically twisted index of ABJM theory\cite{benini}. SUSY-QM also has important applications in mathematical physics, as in providing simple proof of index theorems which establishes connection between topological properties of differentiable manifolds to local properties.

     This review gives a basic introduction to supersymmetric quantum mechanics and later it
establishes SUSY-QM's relevance to the index theorem. We will consider a couple of problems in $0+1$ dimensions, i.e., supersymmetric quantum mechanics, by using supersymmetric path integrals, to illustrate the relationship between physics of the supersymmetric model and geometry of the background space which is some manifold $M$ in the form of Euler characteristic of this manifold $M$. Furthermore, for a manifold admitting spin structure we study a more refined model which yields the index of Dirac operator. Both the Euler characteristic of a manifold $M$ and the index of Dirac operator are the Witten indices of the appropriate supersymmetric quantum mechanical systems. Put differently, we will reveal the connection between supersymmetry and index theorem by path integrals.

     The organization of this paper is as follows: Section II is a an introduction to the calculus of Grassmann variables and their properties. Section III is an introduction to the Gaussian integrals, for both commuting (bosonic) and anticommuting (fermionic) variables including some basic examples. Section IV involves the study of supersymmetric sigma models on both flat and curved space. Section V is the summary and conclusion.

\section{Calculus of Grassmannian variables}

\subsection{Basics}

The use of Grassmannian (non-commutative) variables\cite{supergeometry}\cite{supergeometry1}\cite{supergeometry2}\cite{nakahara}, is indispensable when dealing with supersymmetric theories which is because such theories involve fermionic fields, i.e., these kind of variables are of extreme importance in a superspace formalism. This section will be an elementary introduction to the basic mathematical structure for Grassmann variables, and how the differentiation and integration of such variables is constructed. Such mathematical structure becomes very handy when one requires a classical counterpart of anticommuting operators enroute to developing a quantum theory through quantization scheme.

In compliance with the Bose-Fermi statistics, a collection of Grassmann variables $\{\theta_i\}$ are independent elements of the algebra which anticommute with each other but commute with the ordinary numbers. So, Grassmann variables are anticommuting variables like fermions.

Let $n$ generators $\{\theta_1,\theta_2,...,\theta_n\}$ satisfy the following anticommutation relations

\bea{}
\{\theta_i,\theta_j\}&=&0\nonumber
\\
\Rightarrow \theta_i\theta_j+\theta_j\theta_i&=&0
\label{Grassmann1}
\eea

The above relation implies $\theta_i^2=0$.

Then the set of linear combinations of $\{\theta_i\}$ with the commuting number coefficients is called the Grassmann variable, and with the commuting numbers, $\{\theta_i\}$ satisfy the following relation

\bea{}
\theta_j X^A=X^A\theta_j
\eea

Where $X^A$ are commuting numbers.

Firstly, let us see how a general function of this kind would look like. Let $F(\theta_1,\theta_2,...,\theta_n)$ be a function of Grassmann numbers which satisfy Eq.\eqref{Grassmann1}, each of the $\theta_i$ should appear at most to the power of one. Therefore, a general function of this type could be written in the following form

\bea{}
F=f^0+f^i\theta_i+f^{ij}\theta_i\theta_j+...+f^{12...n}\theta_1\theta_2...\theta_n
\label{Grassmann2}
\eea

where $f$ are real coefficients.

A function of this kind is even when it only has an even number of $\theta_i$ variables in each factor of its expansion \eqref{Grassmann2} and vice versa. However, not every function can be distinguished as even or odd and in those situations one could express such functions as a sum of even and odd functions. Moreover, the following relations also hold

\bea{}
\theta_i^2&=&0 \nonumber
\\
\theta_{i_1}\theta_{i_2}...\theta_{i_n}&=&\epsilon_{i_1i_2...i_n}\theta_1\theta_2...\theta_n\nonumber
\\
\theta_{i_1}\theta_{i_2}...\theta_{i_m}&=&0 \qquad \textmd{if}\quad m>n \nonumber
\eea

where

\bea{}
\epsilon_{i_1i_2...i_n}=\begin{cases}
+1~~~\textmd{if}~i_1i_2...i_n ~\textmd{is}~ \textmd{an}~ \textmd{even}~\textmd{permutation}~ \textmd{of}~1...n \\
-1~~~\textmd{if}~i_1i_2...i_n ~\textmd{is}~ \textmd{an}~ \textmd{odd}~\textmd{permutation}~ \textmd{of}~1...n \\
0 \qquad \textmd{otherwise} \nonumber
\end{cases}
\eea

\subsection{Differentiation}

The right derivatives $\frac{\overrightarrow{\partial}}{\partial\theta_i}$ satisfy

\bea{}
\frac{\overrightarrow{\partial}\theta_i}{\partial\theta_j}&=&\delta_{ij}
\\
\frac{\overrightarrow{\partial}(f\!\cdot\! g)}{\partial\theta_i}&=&(-1)^{|f|}f\frac{\overrightarrow{\partial}g}{\partial\theta_i}+\frac{\overrightarrow{\partial}f}{\partial\theta_i}g
\\
\frac{\overrightarrow{\partial}(af+bg)}{\partial\theta_i}&=&a\frac{\overrightarrow{\partial}f}{\partial\theta_i}+b\frac{\overrightarrow{\partial}g}{\partial\theta_i}
\eea

where $f$ and $g$ are functions, $a$ and $b$ are real or complex numbers and $|f|$ is 1 when $f$ is an odd function of $\theta_i$ and is $0$ when $f$ is an even function of $\theta_i$.

Applying the above properties, differentiation of a general function will be of the following form

\bea{}
\frac{\overrightarrow{\partial}(F)}{\partial\theta_i}=f^i\delta_{ij}f^{jk}\theta_k-\delta_{ij}f^{kj}\theta_k+\delta_{ij}f^{jkl}\theta_k\theta_l-\delta_{ij}f^{kjl}\theta_k\theta_l
+\delta_{ij}f^{klj}\theta_k\theta_l+...
\eea

Left derivative is defined as

\bea{}
\frac{\overrightarrow{\partial}}{\partial\theta_j}\theta_i=\theta_i\frac{\overleftarrow{\partial}}{\partial\theta_j}
\eea

It is just a matter of convenience to choose either left or right derivative.

\subsection{Integration}

Integration can also be defined over these variables, Grassmannian integration is very useful for supersymmetric localization which will become clear later. These kind of integrals satisfy the following rules

\bea{}
\int\left[af(\theta_i)+b~g(\theta_i) \right]d\theta_i&=&a\int f(\theta_i)d\theta_i+b\int g(\theta_i)d\theta_i
\\
\int d\theta_i&=&0
\label{Int2}
\\
\int \theta_id\theta_j&=&\delta_{ij}
\eea
and
\bea{}
\int F_1(\theta_1)F_2(\theta_2)...F_n(\theta_n)d\theta_1d\theta_2...d\theta_n=\int F_1(\theta_1)d\theta_1\int F_2(\theta_2)d\theta_2...\int F_n(\theta_n)d\theta_n
\eea

One interesting feature of Grassmann variables which one can figure out by looking at the properties mentioned above is that the integration for these variables is the same as their differentiation.

Let us apply these rules to the general expression \eqref{Grassmann2}

\bea{}
\int Fd\theta_1d\theta_2...d\theta_n&=&\int(f^0+f^i\theta_i+f^{ij}\theta_i\theta_j+...+f^{12...n}\theta_1\theta_2...\theta_n)d\theta_1d\theta_2...d\theta_n \nonumber
\\
&=&f^{12...n}
\eea

The factor $f^{12...n}$ is the only factor which will survive the integration which is because it is the only factor containing all $\theta_i$s.

The equivalence of differentiation and integration leads to an odd behavior of integration under the change of variables. Let us consider the case of $n=1$ first. Under the change of variables $\theta~\!'=a\theta$, we get the following

\bea{}
\int d\theta f(\theta)&=&\frac{\partial f(\theta)}{\partial\theta} \nonumber
\\
&=&\frac{\partial f\!\left(\frac{\theta'}{a}\right)}{\partial\left(\frac{\theta'}{a}\right)} \nonumber
\\
&=&a\int d\theta'f\left(\frac{\theta'}{a}\right)
\eea

This leads to $d\theta'=\frac{1}{a}d\theta$

This can be extended to the case of $n$ variables. Suppose $\theta_i\rightarrow \theta'_i=a_{ij}\theta_j$, then we have

\bea{}
\int d\theta_1d\theta_2...d\theta_nf(\theta)&=&\frac{\partial}{\partial\theta_1}\frac{\partial}{\partial\theta_2}...\frac{\partial}{\partial\theta_n}f(\theta)\nonumber
\\
&=&\sum\limits_{i_{j=1}}^n \frac{\partial\theta'_{i_1}}{\partial\theta_1}\frac{\partial\theta'_{i_2}}{\partial\theta_2}...\frac{\partial\theta'_{i_n}}{\partial\theta_n}\frac{\partial}
{\partial\theta'_{i_1}}\frac{\partial}{\partial\theta'_{i_2}}...\frac{\partial}{\partial\theta'_{i_n}}f\!\left(\frac{\theta'}{a} \right)\nonumber
\\
&=&\sum\limits_{i_{j=1}}^n \epsilon_{{i_1}{i_2}...{i_n}}a_{i_11}a_{i_22}...a_{i_nn}\frac{\partial}{\partial\theta'_{i_1}}\frac{\partial}{\partial\theta'_{i_2}}...
\frac{\partial}{\partial\theta'_{i_n}}f\!\left(\frac{\theta'}{a} \right)\nonumber
\\
&=&\textmd{det} ~\!a\int d\theta'_1d\theta'_2...d\theta'_n~f\!\left(\frac{\theta'}{a} \right)
\eea

\section{Gaussian Integrals}

Gaussian integrals play an important role in quantum mechanics, quantum field theory, statistical physics etc. These will be quite handy when dealing with sigma model calculations using path integrals later. Therefore it is worthwhile to recall some properties of Gaussian integrals.

%
%
The case of commuting and noncommuting variables will be considered.

\subsection{For commuting variables}

The $n$-dimensional Gaussian integral for commuting variables has the following form

\bea{}
G(\textbf{A})=\int d^nx ~\textmd{exp}\left(-\sum\limits_{i,j=1}^n\frac{1}{2}x_iA_{ij}x_j \right)
\label{G1}
\eea

The above integral \eqref{G1}, converges if the matrix $\textbf{A}$ with the entries $A_{ij}$ is a \textit{symmetric positive definite} matrix.
Different methods give the following result for \eqref{G1}

\bea{}
G(\textbf{A})=(2\pi)^{\frac{n}{2}}(\textmd{det} \textbf{A})^{-\frac{1}{2}}
\label{R1}
\eea

When the matrix is complex, the meaning of square root and determinant need special care.

We should keep in mind that the variables $x$ in all the integrals considered above are commuting real variables. For the commuting complex variables $z$

\bea{}
G(\textbf{A})=\int dz_1d\bar{z}_1...dz_nd\bar{z}_n ~\textmd{exp} \left(-\sum\limits_{i,j=1}^n\frac{1}{2}z_iA_{ij}\bar{z}_j \right)
\label{complex}
\eea

The matrix $A_{ij}$ is $n\times n$ hermitian, and positive definite. The integral \eqref{complex} gives

\bea{}
G(\textbf{A})=(2\pi)^n(\textmd{det} \textbf{A})^{-1}
\label{answer}
\eea

The disappearance of square root sign from \eqref{answer} is not surprising because the integration now runs twice as before.

\subsection{For anticommuting variables}

Let us now consider a Gaussian integral over $n$ Grassmann variables

\bea{}
G(\textbf{A})=\int d\theta_1...d\theta_{2n}~ \textmd{exp}^{\frac{1}{2}\theta_iA_{ij}\theta_j}
\label{Gr1}
\eea

 where $\textbf{A}$ is a real antisymmetric matrix and $\theta$ is a column vector with components $\theta_1,\theta_2,...,\theta_{2n}$. For these integrals, we can not take $\textbf{A}$ to be a symmetric matrix because if we do so, the property \eqref{Grassmann1} of Grassmann variables would immediately imply that the integral \eqref{Gr1} is zero.

 Moreover, each nonzero term in the expansion of the exponential in \eqref{Gr1} contains an even number of $\theta_is$ which must all be different because of the property \eqref{Grassmann1} of Grassmann variables. We must take the integration measure to run upto $2n$ instead of $n$, to make sure that it remains even. If we take it to be $n$, and if $n$ is odd, there is an odd number of factors $d\theta_i$. So there must be atleast one integral $\int d\theta_i$ where the integrand is 1, then \eqref{Int2} implies that \eqref{Gr1} is zero for odd $n$.

 When $n$ is even, the only term which survives in the expansion of the exponential in \eqref{Gr1} is the term which involves $n$ factors of $\theta$. Terms with more than $n$ factors of $\theta$ are immediately zero because of \eqref{Grassmann1} and terms with less than $n$ factors of $\theta$ give zero upon integration because there is atleast one factor $\int d\theta_i$ where the integrand is 1. Thus for $2n$

 \bea{}
 G(\textbf{A})=\frac{1}{2^n~n!}\sum\limits_{\textmd{perms. of}~ i_1...i_{2n}}\epsilon_{i_1...i_{2n}}A_{i_1i_2}A_{i_3i_4}...A_{i_{2n-1}i_{2n}}
 \label{Gr2}
 \eea

 where $\epsilon=\pm1$ is the signature of permutation and the quantity on the right hand side of \eqref{Gr2} is the \textit{Pfaffian} of the antisymmetric matrix \textbf{A}. Therefore

 \bea{}
 G(\textbf{A})=\textmd{Pf}(\textbf{A})
 \eea
 and
 \bea{}
 \textmd{Pf}^2(\textbf{A})=\textmd{det}\textbf{A}
 \eea

%
%
%
%
%
%
%
%
%
%

 Let us consider another integral for $n$ complex Grassmann variables

 \bea{}
 G(\textbf{A})=\int\prod\ d \bar{\theta_i}d\theta_i~\textmd{exp}^{-\bar{\theta_i}A_{ij}\theta_j}
 \eea

 where $\textbf{A}$ is now a hermitian matrix $\bar{A_{ij}}=A_{ji}$ and $\bar{\theta_i}$ are complex conjugates of $\theta_i$. Then by expanding the exponential and using similar arguments as before, we get

 \bea{}
 G(\textbf{A})&=&\int d\bar{\theta_1}d\theta_1...d \bar{\theta_n}d\theta_n \frac{1}{n!}\left[-\bar{\theta}_{i_1}A_{i_1j_1}\theta_{j_1} \right]\left[-\bar{\theta}_{i_2}A_{i_2j_2}\theta_{j_2} \right]...\left[-\bar{\theta}_{i_n}A_{i_nj_n}\theta_{j_n} \right]\nonumber
 \\
 &=&\frac{1}{n!}\int d\bar{\theta_1}...d\bar{\theta_n}\bar{\theta}_{i_1}...\bar{\theta}_{i_n}\int d\theta_1...d\theta_n\theta_{j_1}...\theta_{j_n}
 A_{i_1j_1}...A_{i_nj_n}\nonumber
 \\
 &=& \frac{1}{n!}\epsilon_{i_1...i_n}\epsilon_{j_1...j_n}A_{i_1j_1}...A_{i_nj_n}\nonumber
 \\
 &=&\textmd{det}A
 \eea

 In the above deduction we get a factor $(-1)^{n(n-1)+n(n+1)}=(-1)^{2n^2}=+1$.

%
%
%
%

 \section{Supersymmetric Sigma Models}

 Any sigma model is a geometrical theory of maps from one space to another, for the description of certain physical quantities. Based on different models, such spaces come with extra geometrical structure. In the context of our current study, these spaces will be manifolds. A supersymmetric sigma model contains both bosonic and fermionic fields~\cite{wasay}.

 \subsection{Flat space}

 We first consider a simple model where the target space is flat. Doing the path integral will yield a topological invariant of the underlying space. The Lagrangian of the model is the following

 \bea{}
 \mathcal{L}=\dot x^2+\bar\psi\partial_t\psi
 \eea
which is invariant under the following SUSY transformations

\bea{}
\delta x^i&=&\epsilon\psi^i
\\
\delta\psi^i&=&-\epsilon\dot x^i
\eea
We choose periodic boundary conditions for both bosonic ($x$) and fermionic ($\psi$) fields

\bea{}
x(t+2\pi\beta)=x(t)
\\
\psi(t+2\pi\beta)=\psi(t)
\\
\bar\psi(t+2\pi\beta)=\bar\psi(t)
\eea

The action is given by

\bea{}
S=\int\limits_0^\beta \mathcal{L}dt
\eea

The Witten index for this simple case can be written as a path integral

\bea{}
\textmd{Tr}(-1)^Fe^{-\beta H}=\int\limits_{\textmd{PBCs}}\!\mathcal{D}x\mathcal{D}\psi\mathcal{D}\bar\psi~\textmd{exp}^{-\int\left[\dot x^2+\bar\psi\partial_t\psi\right]dt}
\label{flatspaceaction}
\eea

In the operator $(-1)^F$, $F$ is the fermion number operator. Using integration by parts, we can write the following

\bea{}
\int \dot x^2=-\int x~\partial_t^2~ x\nonumber
\eea

Using this in \eqref{flatspaceaction}, we get

\bea{}
\textmd{Tr}(-1)^Fe^{-\beta H}&=&\int\!\mathcal{D}x\mathcal{D}\psi\mathcal{D}\bar\psi~\textmd{exp}^{-\int\left[x\partial_t^2 x+\bar\psi\partial_t\psi\right]dt}\nonumber
\\
&=&\int\mathcal{D}x~\textmd{exp}^{-\int\left[x\partial_t^2 x \right]dt}~\mathcal{D}\psi\mathcal{D}\bar\psi~\textmd{exp}^{-\int\left[\bar\psi\partial_t\psi \right]dt}
\eea

Now using the integral computations from previous section, we can write

\bea{}
\textmd{Tr}(-1)^Fe^{-\beta H}=\frac{\textmd{det}~\partial_t}{\sqrt{\textmd{det}~ \partial_t^2}}
\eea

The determinant of an operator by properly regularized infinite product of its eigenvalues

\bea{}
\textmd{det}~\partial_t=\prod_{n=1}^\infty\lambda_n^2\nonumber
\eea

Therefore, we can formally write

\bea{}
\sqrt{\textmd{det}~ \partial_t^2}=\textmd{det}~\partial_t\nonumber
\eea
\bea{}
\Rightarrow \textmd{Tr}(-1)^Fe^{-\beta H}=1
\eea

Which is the topological invariant for a flat surface.

\subsection{Curved space}

We now turn our attention towards supersymmetric sigma models with more interesting target space, i.e., a $d$-dimensional Riemannian manifold $M$ with curvature and having nontrivial topology and metric. Since we will not consider deformations by a potential term, we can assume our manifold $M$ to be compact and oriented.

Studying models on curved manifold by evaluating the functional integral for the Witten index of an appropriate supersymmetric quantum mechanical system, one can reveal important connection between physics of the sigma model and geometry of the underlying manifold in the form of characteristic classes~\cite{gaume3}\cite{windey}\cite{windey1}. We will review two such examples, one is the Euler characteristic and second is the index of a Dirac operator.

\subsubsection{Witten index as the Euler characteristic}

The theory involves bosonic variables $x$ which define maps $x\!:S^1\rightarrow M$, where $S^1$ is a one dimensional manifold parameterized by the time $t$, and fermionic variables $\psi$ and $\bar\psi$, which are complex conjugates of each other and odd counterparts of the bosonic variables, are sections of the tangent bundle over $M$: $\psi,\bar\psi\in\Gamma(S^1_\beta,~x^*TM\otimes\mathbb{C})$. $S^1_\beta$ represent a circle of radius $\beta$.

The Lagrangian of this model is

\bea{}
 L=\frac{1}{2}\partial_tx^\mu g_{\mu\nu}\partial_tx^\nu+\frac{i}{2}\bar\psi^\mu \gamma^0 D_t\psi^\nu g_{\mu\nu}+\frac{1}{12}R_{\mu\nu\rho\sigma}(x)\psi^\mu\psi^\nu\bar\psi^\rho\bar\psi^\sigma,
 \label{lag}
\eea

where $g_{\mu\nu}$ is the metric on the manifold $M$, $\psi$ is a two-component real spinor, $R_{\mu\nu\rho\sigma}$ is the Riemann curvature tensor.

We can further simplify Eq.\eqref{lag} in a basis where $\gamma^0$ is diagonal\cite{gaume3},

\bea{}
 L=\frac{1}{2}\partial_tx^\mu g_{\mu\nu}\partial_tx^\nu+i\bar\psi^\mu D_t\psi^\nu g_{\mu\nu}-
 \frac{1}{4}R_{\mu\nu\rho\sigma}(x)\psi^\mu\psi^\nu\bar\psi^\rho\bar\psi^\sigma
 \label{lag2}
 \eea

$D_t$ is the covariant derivative given by

\bea{}
 D_t\psi^\nu=\frac{\partial\psi^\nu}{\partial t}+\dot{x}^\lambda\Gamma^\nu_{\lambda k}(x)\psi^k
 \label{covariantderivative}
\eea

here, $\Gamma$ is the Christoffel symbol of Levi-Civita connection, which preserves the metric and has a vanishing torsion.

The action reads

\bea{}
 S=\int\limits_0^\beta \left[\frac{1}{2}\partial_tx^\mu g_{\mu\nu}\partial_tx^\nu+i\bar\psi^\mu D_t\psi^\nu g_{\mu\nu}-
 \frac{1}{4}R_{\mu\nu\rho\sigma}(x)\psi^\mu\psi^\nu\bar\psi^\rho\bar\psi^\sigma\right]dt
 \label{action}
 \eea

Above action is invariant under the following supersymmetry transformations,
 \bea{}
 \delta x^\nu&=&\epsilon\bar\psi^\nu-\bar\epsilon\psi^\nu
 \\
 \delta\psi^\nu&=&\epsilon(i\dot{x}^\nu-\Gamma^\nu_{\lambda k}\bar\psi^\lambda\psi^k)
 \\
 \delta\bar\psi^\nu&=&\bar\epsilon(-i\dot{x}^\nu-\Gamma^\nu_{\lambda k}\bar\psi^\lambda\psi^k)
 \eea

 where, $\epsilon$ and $\bar\epsilon$ are infinitesimal real and complex Grassmann constants respectively.
 In order to do the path integral computation of the Witten index, we can expand the metric in a Taylor series as follows,

 \bea{}
 g_{\mu\nu}(x_0+\delta x)=g_{\mu\nu}(x_0)+\partial_\alpha g_{\mu\nu}(x_0)\delta x^\alpha+
 \frac{1}{2}\{\partial_\alpha\partial_\beta g_{\mu\nu}(x_0)\delta x^\alpha\delta x^\beta\}+...
 \label{metricexpansion}
 \eea
We can also expand the bosonic and fermionic fields in their relevant fourier modes, as
 \bea{}
 x=x_0+\sum\limits_{n\neq0}\sqrt{\beta}~x_n~e^{\frac{2i\pi nt}{\beta}}
 \\
 \bar\psi=\bar\psi_0+\sum\limits_{n\neq0}\sqrt{\beta}~\bar\psi_n~e^{\frac{2i\pi nt}{\beta}}
 \\
 \psi=\psi_0+\sum\limits_{n\neq0}\sqrt{\beta}~\psi_n~e^{\frac{2i\pi nt}{\beta}}
 \eea

 Note that the Witten index is independent of $\beta$. Also, we can rescale the time as $t\rightarrow\beta t$ in the path integral. Using the above mode expansions, together with Eq. \eqref{metricexpansion}, in Eq. \eqref{action}, we will get

 \bea{}
 S=\sum\limits_{n\neq0}\left[\frac{1}{2}(2in\pi)^2x_n^\mu g_{\mu\nu}(x_0)x_n^\nu\!-\!(2n\pi)\bar\psi^\mu_ng_{\mu\nu}(x_0)\psi^\nu_n-
 \frac{1}{4}R_{\mu\nu\rho\sigma}(x_0)\psi_0^\mu\psi_0^\nu\bar\psi_0^\rho\bar\psi_0^\sigma\right]
 \label{action1}
 \eea

 As mentioned above, the Witten index is independent of $\beta$, therefore we can take the small $\beta$ limit, in particular we can take the limit $\beta\rightarrow 0$. In doing so, we can throw away the terms of the order of $\beta^{\frac{1}{2}}$ and higher order in $\beta$. We have used this in deriving \eqref{action1}.

 Now the path integral approach can be employed

 \bea{}
 I=\int_M \mathcal Dx~\mathcal D\psi~\mathcal D\bar\psi~ e^{-S}=\textmd{Tr}(-1)^Fe^{-\beta H}
 \label{path}
 \eea

with

\bea{}
\mathcal Dx=\prod_n d^dx_n,\mathcal D\psi\!=\!\prod_n d^d\psi_n~ \textmd{and}~ \mathcal D\bar\psi\!=\!\prod_n d^d\bar\psi_n
\eea

where, $d$ is the dimension of manifold, and the integral \eqref{path} will be carried out over the manifold $M$.
The first part of the path integral over the bosonic variables which is corresponding to $\mathcal Dx$, after a rescaling of $x$ will become

\bea{}
I_1=\int\frac{d^dy_n}{2^{\frac{d}{2}}(in\pi)^d} e^{-y^\mu_ng_{\mu\nu}(x_0)y^\nu_n}
\eea

which will give

\bea{}
I_1=\frac{1}{2^{\frac{d}{2}}(in\pi)^d\sqrt{\textmd{det}~g}}
\label{firstpart}
\eea

The second part corresponding to the fermions after a rescaling $\sqrt{2n\pi}~\psi\rightarrow\xi$ can be written as

\bea{}
I_2=(2n\pi)^d\int e^{\bar\xi_ng_{\mu\nu}\xi_n}d^d\xi_nd^d\bar\xi_n
\eea

The metric $g_{\mu\nu}$ is not antisymmetric, so one can make it antisymmetric as follows.
Let $\lambda=\begin{pmatrix}\xi \\\bar\xi\end{pmatrix}$ and $G_{\mu\nu}=\begin{pmatrix}0~~~~ g_{\mu\nu} \\-g_{\mu\nu}^T~~~ 0\end{pmatrix}$, we can rewrite $I_2$ as
\bea{}
I_2&=&(2n\pi)^d\int d^{2d}\lambda ~~e^{\lambda G_{\mu\nu}\lambda}
\\
\Rightarrow I_2&=&\frac{(2n\pi)^d}{2d~!}\int d^{2d}\lambda (\lambda G_{\mu\nu}\lambda)^{2d}\nonumber
\\
&=& \frac{(2n\pi)^d}{2d~!}\epsilon_{i_2}...\epsilon_{i_{2d}}\epsilon_{j_2}...\epsilon_{j_{2d}}G_{\mu_2\nu_2}G_{\mu_{2d}\nu_{2d}}\nonumber
\eea

yields

\bea{}
I_2\sim\frac{(2n\pi)^d}{2d~!}~\sqrt{\textmd{det}G}
\eea

which implies

\bea{}
I_2=\frac{(2n\pi)^d}{2d~!}~\textmd{det}~g
\label{secondpart}
\eea

Now, for the curvature term, which only involves the fermionic zero modes, we have the following integral,
\bea{}
I_3&=&\int\mathcal D\psi\mathcal D\bar\psi~ e^{\frac{1}{4}R_{\mu\nu\rho\sigma}(x_0)\psi_0^\mu\psi_0^\nu\bar\psi_0^\rho\bar\psi_0^\sigma}\nonumber
\\
 &=&\frac{1}{4^{\frac{d}{2}}(\frac{d}{2})!}\int d^d\psi_0~d^d\bar\psi_0~\left(R_{\mu\nu\rho\sigma}(x_0)\psi_0^\mu\psi_0^\nu\bar\psi_0^\rho\bar\psi_0^\sigma \right)^{\frac{d}{2}}\nonumber
\eea

which after performing the integral over the manifold gives
\bea{}
I_3=\frac{1}{4^{\frac{d}{2}}(\frac{d}{2})!}R_{\mu_1\nu_2\rho_1\sigma_2}...R_{\mu_{d-1}\nu_d\rho_{d-1}\sigma_d}\times
\epsilon^{\mu_1\nu_2...\mu_{d-1}\nu_d}\epsilon^{\rho_1\sigma_2...\rho_{d-1}\sigma_d}
\label{thirdpart}
\eea

Combining \eqref{firstpart},\eqref{secondpart} and \eqref{thirdpart} and noting that we have carried out the integral over the manifold for fermionic zero modes, and thus we are left with the integral for the bosonic zero modes only, we have

\bea{}
I=\int d^dx_0 \sqrt{\textmd{det}~g}~ \frac{(-1)^{\frac{d}{2}}}{(4\pi)^{\frac{d}{2}}(\frac{d}{2})!}R_{\mu_1\nu_2\rho_1\sigma_2}...R_{\mu_{d-1}\nu_d\rho_{d-1}\sigma_d}\times
\epsilon^{\mu_1\nu_2...\mu_{d-1}\nu_d}\epsilon^{\rho_1\sigma_2...\rho_{d-1}\sigma_d}
\label{final}
\eea

The Euler class $e(M)$ of a manifold $M$ of dimension $d$ is defined as
\bea{}
e(M)&=&\textmd{pf}\left(\frac{R}{2\pi}\right)\nonumber
\\
&=&\frac{(-1)^{\frac{d}{2}}}{(4\pi)^{\frac{d}{2}}(\frac{d}{2})!}R_{\mu_1\nu_2\rho_1\sigma_2}...R_{\mu_{d-1}\nu_d\rho_{d-1}\sigma_d}\times
\epsilon^{\mu_1\nu_2...\mu_{d-1}\nu_d}\epsilon^{\rho_1\sigma_2...\rho_{d-1}\sigma_d}
\eea

Therefore, the integral in \eqref{final}, can be identified with the Euler number of the manifold $M$, which is also a topological invariant of the manifold $M$ in question and therefore \eqref{final} can be cast into the following form
\bea{}
I=\int_M d^dx_0\sqrt{\textmd{det}~g}~e(M)=\chi(M)
\eea
where, $\chi(M)$ is the Euler number of the manifold $M$, or equally, the Witten index. Therefore we can see that by using the fact that the Wittern index $\textmd{Tr}(-1)^Fe^{-\beta H}$ is independent of $\beta$, we can get the Euler number of the manifold over which the index is defined in terms of a path integral. Moreover, Euler number $\chi(M)$ is a topological invariant of $M$. Thus we can conclude that a physical sigma model defined on a manifold can reveal important topological information about the manifold.

\subsubsection{Witten index as index of a Dirac operator}

We can compute the index of Dirac operator~\cite{atiyah1}\cite{atiyah2}\cite{atiyah3}\cite{atiyah4}, on a $2d$-dimensional spin manifold $M$, by a path integral. We can restrict the model in \eqref{lag} by making the identification~\cite{bagger}, $\psi_1^\mu=\psi_2^\mu=\psi^\mu/\sqrt{2}$, the curvature term vanishes under this restriction and supersymmetries reduce to half. The Lagrangian reduces to

\bea{}
L=\dot x^\mu g_{\mu\nu} \dot x^\nu+g_{\mu\nu}\psi^\mu\frac{D}{Dt}\psi^\nu
\label{lag3}
\eea

The path integral expression for the index of Dirac operator is

\bea{}
\textmd{Ind}~ D=\textmd{Tr}(-1)^Fe^{-\beta H}=\int\limits_{\textmd{PBC}} \mathcal{D}x \mathcal{D}\psi ~\textmd{exp}\!^{-\int\limits_0^\beta dt L}
\eea

The components of the metric in \eqref{lag3} are continuous and differentiable and we can expand the metric in a Taylor series around the point $x_0$

\bea{}
g_{\mu\nu}(x_0+x)=g_{\mu\nu}(x_0)+g_{\mu\nu,\alpha}(x_0)x^\alpha+\frac{1}{2!}~g_{\mu\nu,\alpha\beta}(x_0)x^\alpha x^\beta+...
\label{metricexpansion}
\eea

The symbol $g_{\mu\nu,\alpha}$ and $g_{\mu\nu,\alpha\beta}$ stand for the partial derivatives of $g_{\mu\nu}$ with respect to $x^\alpha$ and $x^\alpha x^\beta$ respectively, evaluated at the point $x_0$. For making the computation simple we will employ Riemann normal coordinates, at the origin of these coordinates, the metric expansion has the following properties

\bea{}
g_{\mu\nu}(x_0)=\delta_{\mu\nu}
\\
g_{\mu\nu,\alpha}(x_0)=0
\eea

Therefore, at the origin of Riemann normal coordinates, \eqref{metricexpansion} becomes

\bea{}
g_{\mu\nu}(x_0+x)=\delta_{\mu\nu}+\frac{1}{2!}~g_{\mu\nu,\alpha\beta}x^\alpha x^\beta+...
\eea

The Lagrangian \eqref{lag3} becomes

\bea{}
L=\left(\delta_{\mu\nu}+\frac{1}{2!}~g_{\mu\nu,\rho\sigma}x^\rho x^\sigma+... \right)\dot x^\mu\dot x^\nu+g_{\mu\nu}\psi^\mu\left(\frac{\partial\psi^\nu}{\partial t}+\dot x^\lambda\Gamma^\nu_{\lambda k}(x)\psi^k \right)
\label{lag4}
\eea

The Christoffel connection is \eqref{lag4} can also be expanded in Taylor series around the point $x_0$

\bea{}
\Gamma^\nu_{\lambda k}(x_0+x)=\Gamma^\nu_{\lambda k}(x_0)+\Gamma^\nu_{\lambda k,\alpha}(x_0)+...
\label{exp}
\eea

where

\bea{}
\Gamma^\nu_{\lambda k}=\frac{1}{2}~g^{\delta\nu}\left(g_{\delta\lambda,k}+g_{\delta k,\lambda}-g_{\lambda k,\delta} \right)
\label{def}
\eea

Together with the properties of metric at the origin of the Riemann normal coordinates and the above definition \eqref{def}, we can safely say that the first term in \eqref{exp} is zero, because it contains the first derivatives of the metric. Then we are left with the higher order terms in \eqref{exp}. Also note that from the bosonic part of the Lagrangian \eqref{lag4} we will only consider the first two terms because the higher order terms will be negligible when we consider small $\beta$ limit $\beta \rightarrow 0$, which will be discussed shortly.

Plugging \eqref{exp} back into \eqref{lag4}

\bea{}
L=\dot x^\mu \dot x^\mu+g_{\mu\nu}\psi^\mu\left[\frac{\partial\psi^\nu}{\partial t}+\dot x^\lambda\psi^k\Gamma^\nu_{\lambda k,\alpha}(x_0)x^\alpha \right]
\label{lag5}
\eea

Since we are working at the origin of Riemann normal coordinates centered at $x_0$, we can write the second derivative of the metric in terms of Riemann curvature tensor by the following relation

\bea{}
\partial_\alpha\partial_\beta g_{\mu\nu}(x_0)=R_{\mu\alpha\nu\beta}(x_0)
\label{id}
\eea

The second term in \eqref{exp} is

\bea{}
\Gamma^\nu_{\lambda k,\alpha}=\frac{1}{2}~g^{\delta\nu}\left(g_{\delta\lambda,k\alpha}+g_{\delta k,\lambda\alpha}-g_{\lambda k,\delta\alpha} \right)
\label{id2}
\eea

using \eqref{id} in \eqref{id2}

\bea{}
g_{\delta\nu}\Gamma^\nu_{\lambda k,\alpha}=\frac{1}{2}\left(R_{\delta\alpha\lambda k}+R_{\delta\alpha k\lambda}-R_{\lambda\alpha k\delta} \right)
\label{prop}
\eea

The Riemann curvature tensor has the following symmetries

\bea{}
R_{\delta\alpha\lambda k}&=&-R_{\alpha\delta\lambda k}
\\
R_{\delta\alpha\lambda k}&=&-R_{\delta\alpha k\lambda}
\\
R_{\delta\alpha\lambda k}&=&R_{\lambda k\delta\alpha}
\\
R_{\delta\alpha\lambda k}+R_{\delta\lambda k\alpha}+R_{\delta k\alpha\lambda}&=&0
\label{bianchi}
\eea

The above properties imply that the Riemann curvature tensor is symmetric under the exchange of first and second pair of indices and antisymmetric under the exchange of two elements in the first and second pair. The last identity \eqref{bianchi} is the well known first Bianchi identity.

Under these properties, \eqref{prop} becomes

\bea{}
g_{\delta\nu}\Gamma^\nu_{\lambda k,\alpha}&=&-\frac{1}{2}R_{\lambda\alpha k\delta}\nonumber
\\
\Rightarrow \Gamma^\nu_{\lambda k,\alpha}&=&-\frac{1}{2}R_{\lambda\alpha k}^\nu
\label{prop1}
\eea

Using \eqref{prop1} in \eqref{lag5}, we get

\bea{}
L=\dot x^\mu \dot x^\mu+g_{\mu\nu}\psi^\mu\left[\frac{\partial\psi^\nu}{\partial t}+\dot x^\lambda\psi^kx^\alpha\left(-\frac{1}{2}R_{\lambda\alpha k}^\nu(x_0) \right) \right]
\label{lag6}
\eea

The term $g_{\mu\nu}\frac{\partial\psi^\nu}{\partial t}$ in \eqref{lag6} can be simplified when we expand the metric in Taylor series, only the first term in the expansion should be considered because all higher order terms are negligible in the small $\beta$ limit, therefore

\bea{}
L=\dot x^\mu \dot x^\mu+\psi^\mu\frac{\partial\psi^\nu}{\partial t}-\frac{1}{2}\dot x^\lambda\psi^kx^\alpha\psi^\mu R_{\lambda\alpha k\mu}(x_0)
\eea

The action is

\bea{}
S=\int\limits_0^\beta L dt
\eea

under the periodic boundary conditions

\bea{}
x(t+2\pi\beta)=x(t)
\\
\psi(t+2\pi\beta)=\psi(t)
\eea

expanding the fields in Fourier modes

\bea{}
x^\mu(t)&=&x_0^\mu+\beta\sum\limits_{n\neq 0}\xi^\mu_n~\textmd{exp}^{\frac{2\pi int}{\beta}}
\\
\psi^\mu(t)&=&\psi_0^\mu+\sqrt{\beta}\sum\limits_{n\neq 0}\eta^\mu_n~\textmd{exp}^{\frac{2\pi int}{\beta}}
\eea

The complete Lagrangian \eqref{lag4} becomes

\bea{}
L=\left(\dot x^\mu\dot x^\mu+\frac{1}{2!}~g_{\mu\nu,\rho\sigma}(x_0)x^\rho x^\sigma\dot x^\mu\dot x^\nu+... \right)+\left(\psi^\mu\frac{\partial\psi^\mu}{\partial t}+\frac{1}{2!}~g_{\mu\nu,\rho\sigma}(x_0)x^\rho x^\sigma\psi^\mu\frac{\partial\psi^\mu}{\partial t}+... \right)-\nonumber
\\
\frac{1}{2}\dot x^\lambda\psi^kx^\alpha\psi^\mu R_{\lambda\alpha k\mu}(x_0)~~~~~
\eea

A rescaling of time as $t\rightarrow \beta t$ yields

\bea{}
S=\int\limits_0^1 \beta[\frac{1}{\beta^2}\left(\beta^2\dot x^\mu\dot x^\mu \right)+\left(\frac{1}{2!}~g_{\mu\nu,\rho\sigma}(x_0)\{\frac{1}{\beta^2}\beta^4x^\rho x^\sigma\dot x^\mu\dot x^\nu \} \right)+...+\nonumber
\\
\left(\frac{\beta}{\beta}\psi^\mu\frac{\partial\psi^\mu}{\partial t}+\frac{\beta^3}{\beta~ 2!}g_{\mu\nu,\rho\sigma}(x_0)x^\rho x^\sigma\psi^\mu\frac{\partial\psi^\mu}{\partial t}+... \right)-\nonumber
\\
\frac{\beta^2}{2\beta}R_{\lambda\alpha k\mu}(x_0)\dot x^\lambda\psi_0^kx^\alpha\psi_o^\mu ]dt
\label{action2}
\eea

We have only expanded the fermions in the curvature term in their corresponding Fourier modes in \eqref{action2}, and only zero modes for them are taken in order to make the curvature term survive in the small $\beta$ limit. The terms of the order of $\beta^3$ and higher order can be neglected in this limit, therefore we are left with

\bea{}
S=\int\limits_0^1\left(\dot x^\mu\dot x^\mu+\psi^\mu\frac{\partial\psi^\mu}{\partial t}-\frac{1}{2}R_{\lambda\alpha k\mu}(x_0)\psi_0^\mu\psi_0^kx^\alpha\dot x^\lambda \right)dt
\eea

Let us now define a fluctuation in the coordinate system as

\bea{}
x^\mu(t)=x_0^\mu+\xi^\mu(t)
\\
\psi^\mu(t)=\psi_0^\mu+\eta^\mu(t)
\eea

Then the second order expansion of the action is

\bea{}
S_f=\int\limits_0^\beta\left[\dot\xi^\mu\dot\xi^\mu+\eta^\mu\dot\eta^\mu-\frac{1}{2}R_{\lambda\alpha k\mu}(x_0)\psi_0^\mu\psi_0^k\xi^\alpha\dot\xi^\lambda \right]dt
\label{action3}
\eea

In \eqref{action3} the operator associated with the $\xi$ field is

\bea{}
-\delta_{\mu\nu}\frac{d^2}{dt^2}+\tilde{R}_{\lambda\alpha}\frac{d}{dt}
\label{1}
\eea
where
\bea{}
\tilde{R}_{\lambda\alpha}=-\frac{1}{2}R_{\lambda\alpha k\mu}(x_0)\psi_0^\mu\psi_0^k\nonumber
\eea

The operator associated with $\eta$ field is

\bea{}
\delta_{\mu\nu}\frac{d}{dt}
\label{2}
\eea

We can now evaluate the index by doing the following path integral

\bea{}
\textmd{Ind}~ D=\int \mathcal{D}\xi\mathcal{D}\eta~e^{-S}
\eea

We will consider the zero modes $\xi_0^\mu$ and $\eta_0^\mu$ separately in the following integral. Note that the operator \eqref{1} is a bosonic operator while the operator in \eqref{2} is a fermionic operator. For the integration over nonzero modes, we have

\bea{}
\textmd{Ind}~ D&=&\int\prod_{\mu=1}^d d\xi_0^\mu d\eta_0^\mu \frac{\sqrt{\textmd{Det}\left(\delta_{\mu\nu}\frac{d}{dt} \right)}}{\sqrt{\textmd{Det}\left(-\delta_{\mu\nu}\frac{d^2}{dt^2}+\tilde{R}_{\lambda\alpha}(x_0)\frac{d}{dt} \right)}}
\\
&=&\int\prod_{\mu=1}^d d\xi_0^\mu d\eta_0^\mu \frac{1}{\sqrt{\textmd{Det}\left(-\delta_{\mu\nu}\frac{d}{dt}+\tilde{R}_{\lambda\alpha}(x_0)\right)}}
\label{functional}
\eea
where $d$ is the dimension of the manifold. Also note that the determinants above are the result of integration over nonzero modes only.

Now we compute the functional determinant in \eqref{functional}, in which the fermionic variables are only contained in $\tilde{R}_{\lambda\alpha}(x_0)$. For the moment, let us suppose that this part is a commuting number, and we know the Riemann tensor is antisymmetric which means that $\tilde{R}_{\lambda\alpha}=-\tilde{R}_{\alpha\lambda}(x_0)$. Since real skew-symmetric matrices are normal matrices, it is possible to bring every skew-symmetric matrix into a block diagonal form by an orthogonal transformation, so it is possible to block diagonalize $\tilde{R}_{\lambda\alpha}$ in an even dimensional manifold, in the following form

\bea{}
\tilde{R}_{\lambda\alpha}=
\left(
  \begin{array}{cccccc}
    0 & \alpha_1 &  & \cdots &  & 0 \\
    -\alpha_1 & 0 &  &  &  & \vdots \\
    \vdots &  &  & \ddots & 0 & \alpha_n \\
    0 &  & \cdots &  & -\alpha_n & 0 \\
  \end{array}
\right)
\eea

Let us focus on the first block, the operator \eqref{1} is real, so its eigenvalues are complex conjugate pairs. When the operator \eqref{1} is applied to the first block, its determinant is given by

\bea{}
\textmd{det}\left(
     \begin{array}{cc}
       -\frac{d}{dt} & \alpha_1 \\
       -\alpha_1 & -\frac{d}{dt} \\
     \end{array}
   \right)&=&\textmd{Det}\left(\frac{d^2}{dt^2}+\alpha_1^2 \right)\nonumber
   \\
   &=&\prod_{n\neq0}\left[\alpha_1^2-\left(\frac{2n\pi}{\beta} \right)^2 \right]\nonumber
   \\
   &=&\left[\prod_{n>1}\left(\frac{2n\pi}{\beta} \right)^2\prod_{n>1}\{1- \left(\frac{\alpha_1\beta}{2n\pi} \right)^2\} \right]^2\nonumber
   \\
   &=&\left[\frac{\textmd{sin}\beta\alpha_1/2}{\alpha_1/2} \right]^2
\eea

If we consider all the $n$ blocks, we will get the following result for \eqref{functional}

\bea{}
\textmd{Ind}~D=\int\prod_{\mu=1}^{2n} d\xi_0^\mu d\eta_0^\mu\prod_{j=1}^n\frac{\alpha_j/2}{\textmd{sin}\beta\alpha_j/2}
\eea

The integration over $\xi_0$ and $\eta_0$ is equivalent to integration of $x_0$ and $\psi_0$ respectively. The product $j$ can be expressed in terms of Riemann curvature tensor, thus yielding

\bea{}
\textmd{Ind}~D=\int\prod_{\mu=1}^{2n}dx_0^\mu d\psi_0^\mu\frac{1}{\beta^{\frac{d}{2}}}\textmd{det}\left(\frac{\beta\tilde{R}/2}{\textmd{sin}\beta\tilde{R}/2} \right)^{\frac{1}{2}}
\eea

We can make the following change of variables to remove $\beta$ dependence

\bea{}
\psi_0^\mu=\frac{\chi_0^\mu}{\sqrt{\beta}}~;~~d\psi_0^\mu=\sqrt{\beta}~d\chi_0^\mu\nonumber
\eea

and substitute

\bea{}
\beta\tilde{R}=-\frac{1}{2}R_{\lambda\alpha k\mu}(x_0)\psi_0^\mu\psi_0^k \nonumber
\eea

Therefore,

\bea{}
\textmd{Ind}~D=\int\prod_{\mu=1}^{2n}dx_0^\mu d\psi_0^\mu~\textmd{det}\left(\frac{\frac{1}{2}\frac{1}{2}R_{\lambda\alpha k\mu}(x_0)\psi_0^\mu\psi_0^k}{\textmd{sin}\frac{1}{2}\frac{1}{2}R_{\lambda\alpha k\mu}(x_0)\psi_0^\mu\psi_0^k} \right)^{\frac{1}{2}}
\label{result}
\eea

Note that we have dropped infinitely many terms while going through the computations, therefore \eqref{result} is a close approximation to the actual result. If we take this into account along with taking care of the factors of $i/2\pi$ arising from the Feynman measure in the integral, then after carrying out the integral over zero modes we can get

\bea{}
\textmd{Ind}~D=\int \textmd{det}\left(\frac{\frac{1}{2}\frac{1}{2\pi}R}{\textmd{sinh}\frac{1}{2}\frac{1}{2\pi}R} \right)^{\frac{1}{2}}
\label{index}
\eea
where the integral is over the manifold $M$.

The relation \eqref{index} is a well known form of the index of a Dirac operator.

\section{summary and conclusion}

In this paper, we started by reviewing some basics about the Grassmann variables. Later in section III we reviewed Gaussian integrals involving both commuting and anticommuting variables. In section IV we started to look into the supersymmetric sigma models, firstly we computed a simple supersymmetric sigma model on a flat space by path integrals, and found that it yields the topological invariant of the flat space which is identified with the Witten index.

The next two models are supersymmetric sigma model defined on a curved space. First one \eqref{lag} is on a general Riemann manifold $M$ and the second one on a spin manifold. We do the computations using path integrals in the classical limit $\beta \rightarrow 0$. In the usual mathematical sense, Path integrals are infinite dimensional, however for the cases considered, the computation reduces to a finite dimensional integration or in other words the path integral localizes to the zero modes, which then enables us to establish a connection with the standard definitions of topological invariants.

In the case of \eqref{lag}, the supersymmetric path integral derivation for the Witten index yields the Euler characteristic of the target manifold $M$, which is the Euler class of the vector bundle $T^\ast M$. The second model \eqref{lag3} on a spin manifold is a obtained from \eqref{lag} by making the identification $\psi_1^\mu=\psi_2^\mu=\psi^\mu/\sqrt{2}$. A similar supersymmetric path integral computation in the semiclassical limit yields the index of Dirac operator.

Finally, one could compute these invariants for the restricted Witten index of the appropriate supersymmetric quantum mechanical systems\cite{restricted}, the work in this direction is in progress.

\section*{Acknowledgments}

The author would like to thank Maxim Zabzine for introducing the subject and guidance during the stay at Uppsala. The author would also like to thank Yasir Jamil for providing the facilities to carry out this work. In addition, the author would like to thank Ming Chen for providing some useful references and comments. Lastly, the author would like to thank the anonymous referee for his helpful comments and suggestions.

\section*{Conflict of Interests}

The author declares that there is no conflict of interests regarding the publication of this paper.


\vspace*{3pt}

\begin{thebibliography}{0}
\bibitem{zumino}
  J. Wess and B. Zumino, Nucl. Phys. B 70 (1974) 39.
   \bibitem{Likhtman}
   Y.A. Gel'fand and E.P. Likhtman, JETP Lett. 13 (1971) 323.
   \bibitem{Ramond}
   P. Ramond, Phys. Rev. D 3 (1971) 2415.
   \bibitem{Schwarz}
   A. Neveu and J. Schwarz, Nucl. Phys. B 31 (1971) 86.
   \bibitem{volkov}
  D. Volkov and V. Akulov, Phys. Lett. B 46 (1973) 109.

  \bibitem{greene}
   B. Greene, The elegant universe, W. W. Norton, 1999.
   \bibitem{mandula}
   S. Coleman and J. Mandula, Phys. Rev. 159 (1967) 1251.
   \bibitem{freedman}
  S. Ferrara, D. Freedman and P. van Nieuwenhuizen, Phys. Rev. D 13 (1976) 3214.
  \bibitem{deser}
  S. Deser and B. Zumino, Phys. Lett. B 62 (1976) 335.
   \bibitem{witten}
  E. Witten, Dynamical breaking of supersymmetry, Nucl. Phys. B 188 (1981) 513.
    \bibitem{cooper}
  F. Cooper and B. Freedman, Ann. Phys. 146 (1983) 262.
  \bibitem{witten1}
  E. Witten, Nucl. Phys. B 202 (1982) 253.
  \bibitem{bender}
   C. Bender, F. Cooper and A. Das, Phys. Rev. D 28 (1983) 1473.
  \bibitem{comtet}
   R. Akhoury and A. Comtet, Nucl. Phys. B 246 (1984) 253.
  \bibitem{khare}
   A. Khare and J. Maharana, Phys. Lett. B 145 ( 1984) 77.
  \bibitem{cecotti}
   S. Cecotti and L. Girardello, Ann. Phys. 145 (1983) 81.
  \bibitem{cooper1}
  B. Freedman and F. Cooper, Physica D 15 (1985) 138.
  \bibitem{Salomonson}
  A. Kihlbexg, P. Salomonson and B.S. Skagerstam, Zeit. Phys. C 28 (1985) 203.
  \bibitem{Gaume}
  L. Alvarez-Gaume, J. Phys. A 16 (1983) 4177.
  \bibitem{Gaume1}
  L. Alvarez-Gaume and E. Witten, Nucl. Phys. B 234 (1984) 269.
  \bibitem{Gaume2}
  D. Friedan and P. Windey, Physica D 15 (1985) 71.
  \bibitem{parisi}
  G. Parisi and N. Sourlas, Nucl. Phys. B 206 (1982) 321.
  \bibitem{holten}
  P. Salomonson and J. van Holten, Nucl. Phys. B 196 (1982) 509.
  \bibitem{prl}
  L.F. Urrutia and E. Hemandez, Phys. Rev. Lett. 51 (1983) 755.
  \bibitem{khare1}
  A. Khare and J. Maharana, Nucl. Phys. B 244 ( 1984) 409.
  \bibitem{prl1}
  A.B. Balantekin, Ann. Phys. 164 (1985) 277.
  \bibitem{prl2}
  A. Kostelecky and M.M. Nieto, Phys. Rev. Lett. 53 (1984) 2285.
  \bibitem{prl3}
  A. R. P Rau, Phys. Rev. Lett. 56 ( 1986) 95.
  \bibitem{stedman}
  C. Blockley and G. Stedman, Eur. J. Phys. 6 (1985) 218.
  \bibitem{gauntlett}
  J.P.Gauntlett, Nucl. Phys. B 411. 443. (1994).
  \bibitem{hollowood}
  T.J.Hollowood and T.Kingaby, Phys. Lett. B 566 (2003) 258.
  \bibitem{pachos}
   J. K. Pachos and M. Stone, Int. J. Mod. Phys. B 21, 5399 (2007).
   \bibitem{witten2}
   E. Witten, J. Diff. Geom. 17 (1982) 661.
   \bibitem{gaume3}
   L. Alvarez-Gaume, Commun. Math. Phys. 90 (1983) 161.
   \bibitem{windey}
   D. Friedan and P. Windey, Nucl. Phys. B 235 (1984) 395.
   \bibitem{windey1}
   P. Windey, Supersymmetric quantum mechanics and the Atiyah-Singer index theorem, Acta Phys. Polon. B 15 (1984) 435.
   \bibitem{witten3}
   E. Witten, Commun. Math. Phys. 121 (1989) 351.
   \bibitem{witten4}
   E. Witten, Commun. Math. Phys. 117 (1988) 353.
   \bibitem{benini}
    F. Benini, K. Hristov, A. Zaffaroni, \url{http://arxiv.org/abs/1511.04085}.
   \bibitem{supergeometry}
   A. Rogers, Supermanifolds: Theory and applications, World Scientific Publishing (2007).
   \bibitem{supergeometry1}
   B. DeWitt, Supermanifolds, Cambridge University Press (1984).
   \bibitem{supergeometry2}
   K. Hori et al., Mirror Symmetry, Clay mathematics monographs (2003).
   \bibitem{nakahara}
   M. Nakahara, Geometry Topology and Physics, 2nd Edition, IOP Publishing Ltd (2003).
   \bibitem{wasay}
   M.A.Wasay, Supersymmetric quantum mechanics, MS. Thesis, Uppsala University (2010).
   \bibitem{atiyah1}
   Atiyah M.F., Singer I.M., The index of elliptic operators. I, Ann. of Math. (2) 87 (1968), 484–530.
   \bibitem{atiyah2}
   Atiyah M.F., Singer I.M., The index of elliptic operators. III, Ann. of Math. (2) 87 (1968), 546–604.
   \bibitem{atiyah3}
   Atiyah M.F., Singer I.M., The index of elliptic operators. IV, Ann. of Math. (2) 93 (1971), 119–138.
   \bibitem{atiyah4}
   Atiyah M.F., Singer I.M., The index of elliptic operators. V, Ann. of Math. (2) 93 (1971), 139–149.
   \bibitem{bagger}
   J. Bagger, Supersymmetric sigma models, In Proceedings of the Bonn-NATO
Advanced Study Institute on Supersymmetry, New York, Plenum, 1985, p. 213-257.
\bibitem{restricted}
M. Campostrini, J. Wosiek, Phys. Lett. B 550 (2002) 121.
\end{thebibliography}
\end{document}